\documentclass[aps,prl,groupedaddress,twocolumn]{revtex4}

\usepackage{amsmath}
\usepackage{graphicx}

\newcommand{\D}{\mathrm{d}}
\newcommand{\Ai}{\mathrm{Ai}}
\newcommand{\tr}{\mathrm{tr}}

\newcommand{\bra}[1]{\langle #1 \vert}
\newcommand{\ket}[1]{\vert #1 \rangle}
\newcommand{\braket}[2]{\langle #1 \vert #2 \rangle}

\begin{document}

\title{Width scaling of an interface constrained by a membrane}

\author{J. Whitehouse}
\author{R. A. Blythe}
\author{M. R. Evans}
\affiliation{SUPA, School of Physics and Astronomy, University of Edinburgh, Peter Guthrie Tait Road, Edinburgh EH9 3FD}

\author{D. Mukamel}
\affiliation{Department of Physics of Complex Systems, Weizmann Institute of Science, Rehovot, Israel 7610001}

\date{26 July 2018}

\begin{abstract}
We investigate the shape of a growing interface in the presence of an impenetrable moving membrane. The two distinct geometrical arrangements of the interface and membrane, obtained by placing the membrane behind or ahead of the interface, are not symmetrically related. On the basis of numerical results and an exact calculation, we argue that these two arrangements represent two distinct universality classes for interfacial growth: whilst the well-established Kardar-Parisi-Zhang (KPZ) growth is obtained in the `ahead' arrangement, we find an arrested KPZ growth with a smaller roughness exponent in the `behind' arrangement. This suggests that the surface properties of growing cell membranes and expanding bacterial colonies, for example, are fundamentally distinct.
\end{abstract}

\maketitle

\noindent{\em Introduction.}---Models of growing interfaces have been pivotal in developing a theoretical understanding of nonequilibrium statistical physics \cite{HHZ,Krug97,KK10,Takeuchi17}. In particular, they have revealed that  concepts of scaling and universality can apply beyond equilibrium critical phenomena to systems driven out of equilibrium. An important, robust universality class of growing interfaces is described by the Kardar-Parisi-Zhang (KPZ) equation, which  reads
\begin{equation}
\frac{\partial h}{\partial t} = D \nabla^2 h
+ \lambda \left( \nabla h \right)^2 + \xi \;,
\label{KPZ}
\end{equation}
where $h(x,t)$ is the height of an interface above a substrate and $\xi(x,t)$ is a Gaussian white noise \cite{KPZ}.  It is well established that the width, $W$, of a growing interface exhibits a dynamical scaling form \cite{FV}
\begin{equation}
W  \sim L^{\alpha} f(t/L^z) \label{width}
\end{equation}
where the roughness exponent $\alpha$ and dynamical exponent $z$ are determined by the universality class of the interface ($\alpha=\frac{1}{2}$ and $z=\frac{3}{2}$ for the KPZ equation in one dimension \cite{KPZ,HHZ,Krug97}). 
Furthermore, universality also applies to higher moments of  the height distribution \cite{KMHH}.

More recently, major progress has been made in showing that  the long time 
distribution of the height follows that of the largest eigenvalue of a random matrix drawn from an ensemble that depends on the geometry of the  interface \cite{Johansson,PS,SS}. Remarkably, these different sub-universality classes have been observed experimentally \cite{TS,TSSS}. Very recently exact results for random initial conditions have been obtained \cite{CFS17,MS17} as well as large deviations for  the tails of the height  distribution \cite{MKV16,LDMRS16}.

\begin{figure}
  \centering
  \includegraphics[width=0.8\linewidth]{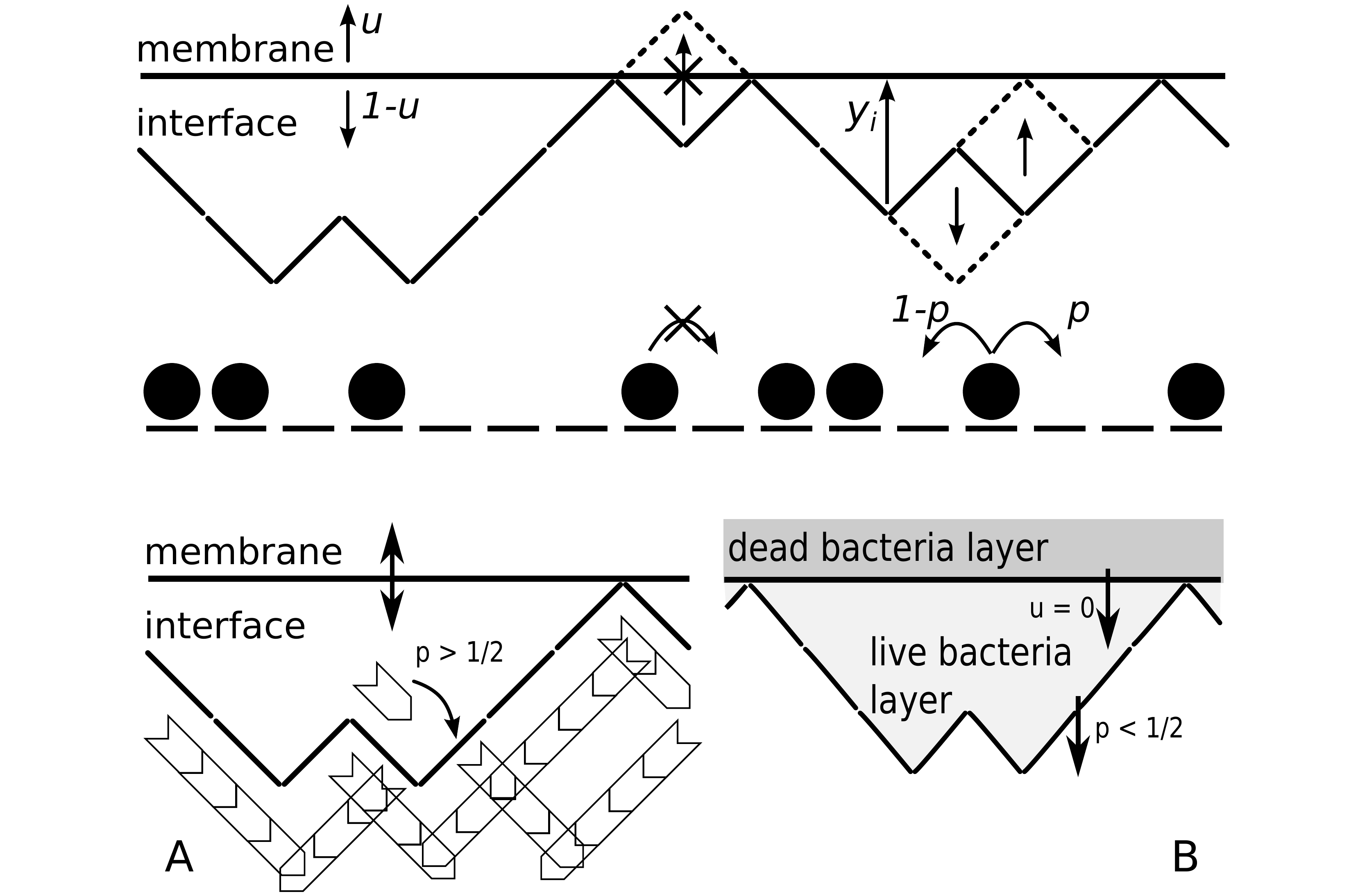}
 \caption{(Upper panel) Membrane-Interface Model and mapping to an exclusion process. (Lower left) Arrangement A: actin network with leading interface interacting with a cell membrane. (Lower right) Arrangement B: bacterial colony growth of live cells at an interface with a dead layer following behind.}
\label{figMIM}
\end{figure}

In this work we consider  how the shape of a growing interface is affected by an impenetrable flat membrane whose vertical position fluctuates (see Fig.~\ref{figMIM}). As we note below, different physical situations correspond to two distinct arrangements: one in which the membrane is \emph{ahead} of the growing interface, and one in which it is \emph{behind}. We refer to these as arrangements A and B respectively. Since the KPZ equation is not invariant under height-reversal  ($h\to-h$), the two arrangements need not belong to the same universality class. On the basis of numerical results, an exact solution along a special line  and crossover scaling analyses, we argue that this system has a rich phase diagram comprising two distinct phases where the interface is rough and bound to the membrane (in addition to an unbound and a smooth, bound phase). Specifically, rough interfaces in arrangement A are described by the usual KPZ scaling exponents, while those in arrangement B have distinct exponents $\alpha=\frac{1}{3}$ and $z=1$. We further argue that the latter behavior arises from  dynamical arrest of the growing interface and therewith a generic mechanism for non-KPZ scaling.

Such deviations from KPZ behavior are of interest because the KPZ universality class is highly robust, due to the nonlinearity $(\nabla h)^2$ being the only relevant contribution to the dynamics of growing interfaces with local interactions at large length- and timescales for $d \leq 2$
\cite{BS,LP}. Known exceptions tend to occur in rather special cases \cite{KS88,MKHZ89,DLSSa,DLSSb,WT90,KM92}, for example, when parameters are tuned to generate a leading nonlinearity at higher-order or when the noise is colored.

\noindent\textit{Model definition.}---We now  define the membrane-interface model that we study (see Fig.~\ref{figMIM}). The interface is of  restricted solid-on-solid type \cite{MRSB86,KK89} and consists of $L$ upward or downward horizontal steps with periodic boundary conditions. We exploit a mapping to a system of particles with hard-core interactions where an occupied site, $\tau_i=1$, represents a down-step of the interface (reading from left to right) and an empty site, $\tau_i=0$, represents an up-step. On average once per unit time, each of the $L/2$ particles attempts to hop to a neighboring site (with the move rejected if the target site is occupied). Once a particle is chosen, it hops to the right with probability $p$ (which corresponds to a valley in the interface rising by two lattice sites) and to the left with probability $1-p$. Free of a constraining membrane, the stationary state is one in which all particle configurations are equally likely \cite{LP,MRSB86}: the interface then has the shape of a random walk, and its centre-of-mass moves upwards with velocity $v_I^f = p - \frac{1}{2}$.

The dynamics of the free interface is modified by the presence of the membrane, a flat horizontal wall that is positioned above the interface (see Fig. 1). Once per unit time, the membrane attempts to move by one lattice site: upwards with probability $u$, and downwards with probability $1-u$. Thus in the absence of the interface, the membrane moves upwards with velocity $v_M^f=2u-1$. The interaction between the interface and membrane is modeled as follows: \emph{Any} move, of either the interface or the membrane, that would cause them to overlap is \emph{forbidden}. This occurs when there are points of contact between the membrane and interface. Note that the `ahead' (A) and `behind' (B) arrangements are both included in the model definition: (A) when $p>\frac{1}{2}$, the interface's growth is directed towards the membrane: i.e., any contacts with the membrane occur at interfacial peaks; (B) when $p<\frac{1}{2}$ the situation is reversed, and contacts lie at interfacial troughs.  Since the height distribution functions of the free interface are asymmetric about the mean position \cite{Johansson,PS,SS}, the effect of the  membrane may be different in the two arrangements.

Arrangements A and B are also motivated by biophysical considerations. In cell growth, fluctuations of the cell membrane can allow a lamellipodium (a sheet of material packed with growing actin filaments) to expand into the space behind it \cite{Bray01,WBFEGM}. At the same time, the membrane inhibits the lamellipodium, and the system can be modeled by arrangement A  (see Fig.~\ref{figMIM}, lower-left panel). By contrast, some bacterial colonies advance through the division of living cells on the interface between the colony and its surroundings \cite{WSG17}. The consumption of nutrients by bacteria in the surface layer starve bacteria behind them: the boundary between the living and dead bacteria then forms a membrane behind the interface, as in arrangement B (Fig.~\ref{figMIM}, lower-right panel; \cite{BWpc}). More generally,  the motion induced by growth  behind a point-like fluctuating barrier has been modeled as a Brownian ratchet \cite{POO93}: the model we introduce here thus 
realizes  a \emph{spatially-extended} ratchet.

\noindent\textit{Phase diagram.}--- We establish the behavior  for 
different values of $p$ and $u$ by direct Monte Carlo simulation. 
We define the separation $y_i\ge0$ as the distance between the interface and membrane at lattice point $i$, and use an initial condition in which the interface lies flat against the membrane (i.e., $y_i$ is alternately $0$ and $1$). In characterizing the interfacial properties, it is important to distinguish between spatial averages, such as the centre-of-mass separation $\overline{y} = \frac{1}{L} \sum_{i=1}^L y_i$, and averages over the dynamical ensemble, which we denote with angle brackets. Then, $\overline{y}$ is a random variable, and the width $W$ is defined by $W = \langle [\overline{(y-\overline{y})^2}]^{\frac{1}{2}} \rangle$. We also consider the number of contacts between the membrane and interface, $C = L \langle \overline{\delta_{y,0}} \rangle$.

\begin{figure}
  \centering
  \includegraphics[width=0.8\linewidth]{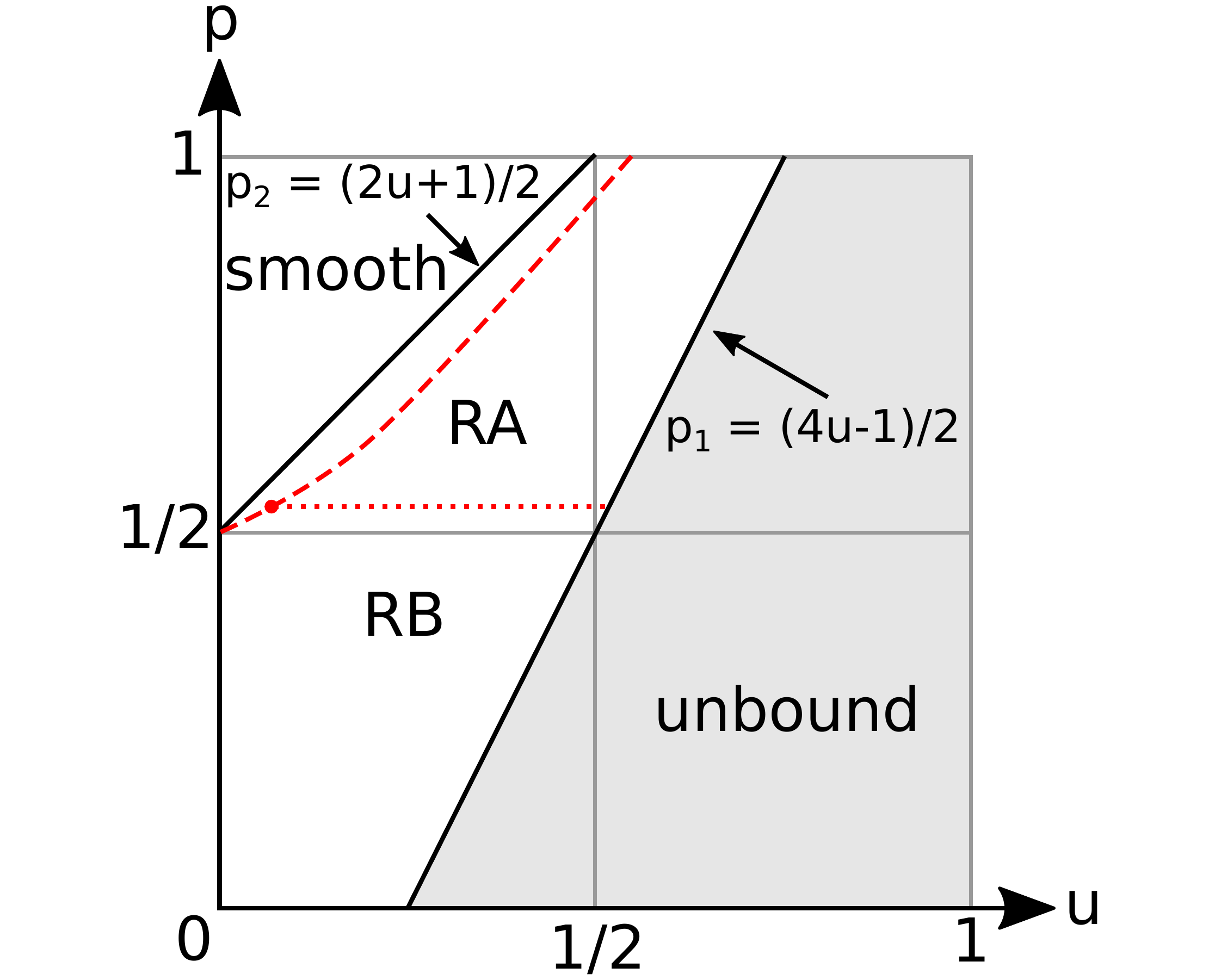}
 \caption{Semi-quantitative phase diagram in $u$--$p$ plane. RA and RB denote two rough phases, found in arrangements A and B, respectively, and in which the roughness exponents are $\alpha=\frac{1}{2}$ and $\alpha=\frac{1}{3}$, respectively. Dotted line indicates finite-size offset of the transition line between RA and RB; dashed line the offset of the rough-smooth transition from the upper bound $p_2=u+\frac{1}{2}$.}
\label{figPD}
\end{figure}

We summarize our findings from the simulations in the form of a phase diagram comprising four phases, Fig.~\ref{figPD}. When $v_M^f = 2u-1 > v_I^f = p-\frac{1}{2}$, a free membrane and interface recede from each other. Thus for $p<p_1 = (4u-1)/2$ we expect an \emph{unbound phase} where $\overline{y}$ increases indefinitely, $C \sim 0$ and the interface has KPZ scaling. Meanwhile, the membrane cannot move faster than $v_M=u$, even when pushed by the interface. Therefore, if $v_I^f = p-\frac{1}{2} > u$, we expect for $p>p_2=(2u+1)/2$ a \emph{smooth} phase where $C \sim O(L)$ and $W\sim O(1)$. Simulation data (see Supplementary Figures) are consistent with these expectations and  suggest that the boundary between the rough and smooth phase lies slightly below $p_2$, as indicated in Fig.~\ref{figPD}. 

\begin{figure}
  \centering
  \includegraphics[width=\linewidth]{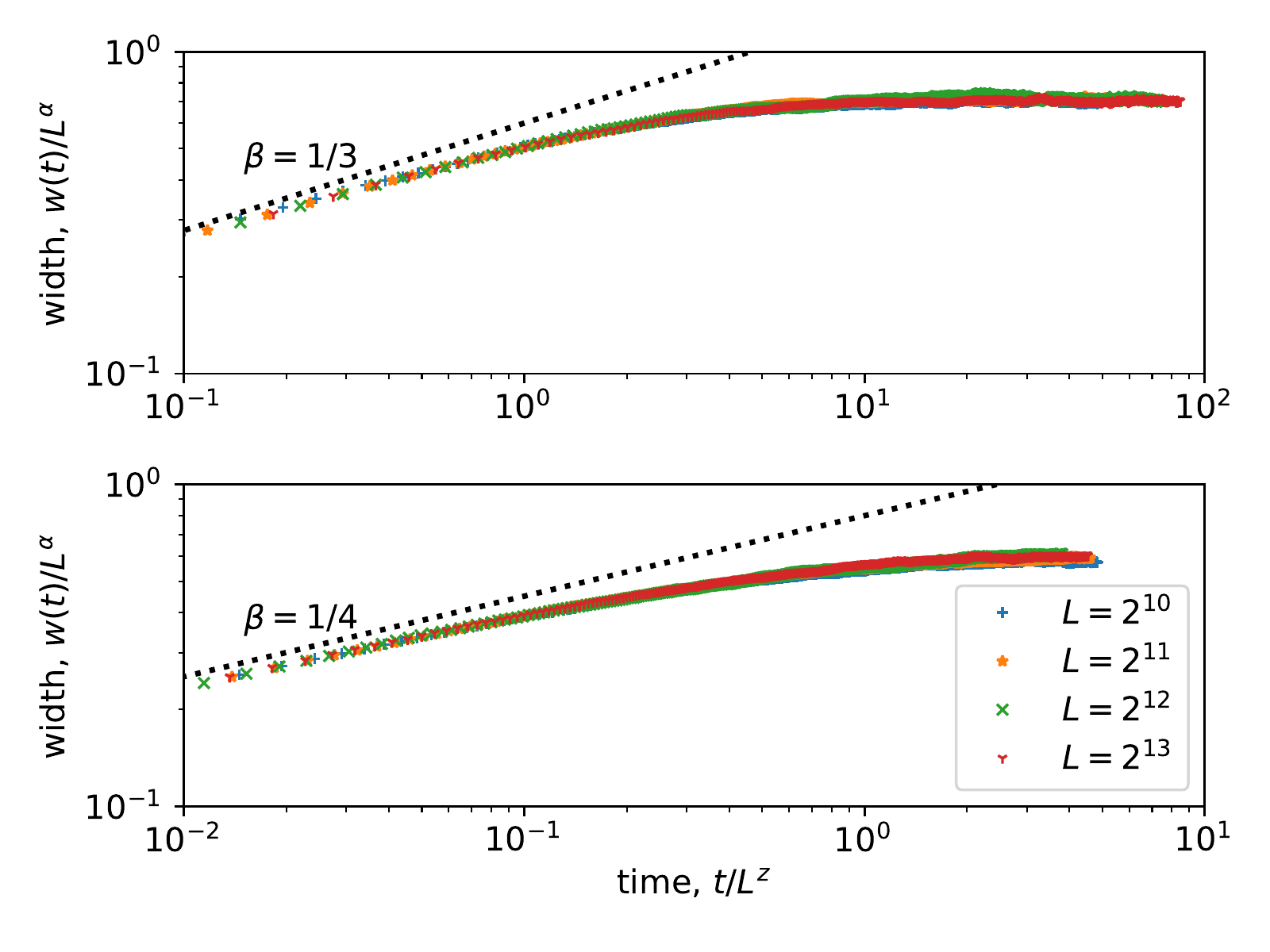}
 \caption{Dynamical scaling collapse. Upper panel: RB phase ($p=0.15, u=0.15$), $\alpha=\frac{1}{3}$, $\beta=\frac{1}{3}$, $z=1$ (arrested KPZ). Lower panel: transition line ($p=0.5, u=0.35$), $\alpha=\frac{1}{3}$, $\beta=\frac{1}{4}$, $z=\frac{4}{3}$ (arrested EW).}
\label{fig:dynscal}
\end{figure}

Of greatest interest  are the two rough phases that lie between $p_1$ and $p_2$, with a transition between them near $p=\frac{1}{2}$. From the upper panel in Fig.~\ref{fig:dynscal} we find that for $p<\frac{1}{2}$ (arrangement B), the interface width scales as $L^\alpha$ with an exponent $\alpha=\frac{1}{3}$, smaller than the usual KPZ exponent of $\alpha=\frac{1}{2}$. By contrast, for $p>\frac{1}{2}$ (arrangement A), $\alpha=\frac{1}{2}$ is obtained (see Supplementary Figures). In both rough phases the contact count is $C\sim O(1)$ i.e. the interface touches the membrane only a few times, with long excursions in between. In the following, we exclude the possibility that the non-KPZ exponent is a finite-size effect (as is sometimes the case \cite{BE2001}) on the basis that an exact calculation, a crossover scaling analysis and a dynamical scaling collapse, all consistently point to $\alpha=\frac{1}{3}$. These analyses also suggest that in the limit $L\to \infty$ the transition between the two rough phases occurs exactly at $p=\frac{1}{2}$, where the switch between arrangements A and B takes place. We thus designate these phases as RA and RB, respectively.

\noindent{\em Exact solution in the weakly-asymmetric limit.}---
We now show that the stationary state of the model is exactly solvable along the line 
\begin{equation}p=\frac{1}{2} + \frac{c(u)}{L}\label{pwa}\end{equation} 
where particle hopping is weakly asymmetric 
($p\to 1/2$ as $L\to \infty$) and $c(u)$ is chosen so that detailed balance is satisfied.    We first  make a simple ansatz for the stationary distribution, $P(\{y_i\}) \propto \prod_i q^{y_i}$, where $q$ is to be determined. For detailed balance to hold for  membrane movement, we require $u P(\{ y_i \}) =(1-u) P(\{ y_i +1 \})$ which yields $u/(1-u) = q^L$. Likewise, for the interface move $y_i \to y_i-2$ we require $p  q^{y_i} =(1-p) q^{y_i-2}$ yielding
$q^2= (1-p)/p$. Combining the two conditions gives
\begin{equation}
u  = \frac{(1-p)^{L/2}}{p^{L/2} + (1-p)^{L/2}}\label{upcon}\;.
\end{equation}
In the limit $L\to \infty$ this corresponds to $c(u) \simeq \frac{1}{2}\ln(\frac{1}{u}-1)$ and $q\simeq  1- 2\frac{c(u)}{L}$. Thus as $L\to \infty$ the exactly solvable line approaches $p=\frac{1}{2}$ from above for all $0\leq u \leq \frac{1}{2}$. This suggests that the dotted line between RA and RB in Fig.~\ref{figPD} is a finite-size effect.

To determine the stationary height distribution $P(y)$, we introduce  basis vectors $\{\ket{0}, \ket{1}, \ldots \}$ that span the space of heights, and  transfer matrix defined by $T\ket{0} = \ket{1}$ and $T\ket{y} = q^y (\ket{y-1}+\ket{y+1})$ for $y>0$. The expression
\begin{equation}
\label{Py}
  P(y) = \frac{\bra{y} T^L \ket{y}}{\tr(T^L)} \simeq \frac{\braket{y}{\phi}\braket{\psi}{y}}{\tr(\ket{\phi}\bra{\psi})}
\end{equation}
builds in the constraint that the height changes by exactly one unit along each bond on the lattice. Here we have used the fact that, for large $L$, $T^L \simeq \ket{\phi}\mu^L\bra{\psi}$ if $\mu$ is the largest eigenvalue of $T$ and $\ket{\phi}$ and $\bra{\psi}$ are the corresponding right and left eigenvectors.

The eigenvalue equation $T\ket{\phi} = \mu \ket{\phi}$ reads
\begin{equation}
\label{recursion}
q^y [(1-\delta_{y,0}) \braket{y-1}{\phi} + \braket{y+1}{\phi}] = \mu \braket{y}{\phi} \;.
\end{equation}
We now make a continuum approximation $\braket{y}{\phi} = \phi(y)$, put $q=1-\frac{\epsilon}{2}$ where $\epsilon=\frac{4c(u)}{L}$, and expand to leading non-trivial order for the case $y>0$. We find
\begin{equation}\label{eq:ctm}
  (2-\mu - y\epsilon) \phi(y) + \frac{\D^2 \phi}{\D y^2} = 0 \;,
\end{equation}
which, under a change of variable
$z = (\mu-2)\epsilon^{-\frac{2}{3}} + \epsilon^{\frac{1}{3}}y$ and $\phi(y) \to f(z)$ transforms to Airy's equation
\begin{equation}\label{eq:Airy-eq}
  \frac{\D^2 f(z)}{\D z^2} - zf(z) = 0 \;.
\end{equation}
The physical solution, which vanishes as $z\to\infty$, is $f(z)=\Ai(z)$. The appearance of the combination $\epsilon^{\frac{1}{3}}y$ is almost sufficient to establish that characteristic lengthscale of the interface scales as $\epsilon^{-\frac{1}{3}} \sim L^{\frac{1}{3}}$, but to confirm this requires knowledge of how the eigenvalue $\mu$ behaves as $\epsilon\to0$. This we deduce from the boundary case $y=0$ in Eq.~(\ref{recursion}), which  becomes
\begin{equation}\label{eq:bc}
  \left( \mu -1 \right) \phi(0) = \left. \frac{\D \phi}{\D y} \right|_{y=0} 
\end{equation}
or equivalently $(\mu-1)f(z^*) = \epsilon^{\frac{1}{3}} f'(z^*)$ where $z^*=(\mu-2)\epsilon^{-\frac{2}{3}}$. This equation is satisfied by $z^* = z_0 + \beta \epsilon^{\frac{1}{3}}$ where $z_0=  -2.3381 \ldots $ is the largest real root of the Airy function. Substituting the resulting expression for $\mu$ back into the boundary condition reveals that $\beta = 1 + O(\epsilon^{\frac{1}{3}})$. Putting this together we find
\begin{equation}\label{eq:phi_y}
  \phi(y) = \Ai \bigg( (y+\beta) \epsilon^{\frac{1}{3}} - |z_0| \bigg)
\end{equation}
and a similar  expression for the left eigenvector  $\psi(y)$. Then, from (\ref{Py}) we finally obtain
\begin{equation}
P(y) \simeq \epsilon^{\frac{1}{3}} \frac{ \Ai^2 \bigg( (y+\beta) \epsilon^{\frac{1}{3}} - |z_0| \bigg)
}{\int_{z_0}^\infty \Ai^2 (u) \D u}
\label{P(y)}
\end{equation}
 which readily implies that $\langle \overline{y^n} \rangle \sim \epsilon^{-\frac{n}{3}} \sim L^{\frac{n}{3}}$, as $\epsilon\sim \frac{1}{L}$. In particular, the width $W\sim L^{\frac{1}{3}}$, confirming the roughness exponent $\alpha$ in the vicinity of the line $p=\frac{1}{2}$. In addition, $P(0) \sim \epsilon$ which is consistent with a mean number of contacts that is independent of $L$: $C= L P(0) = O(1)$.

\begin{figure}
  \centering
  \includegraphics[width=\linewidth]{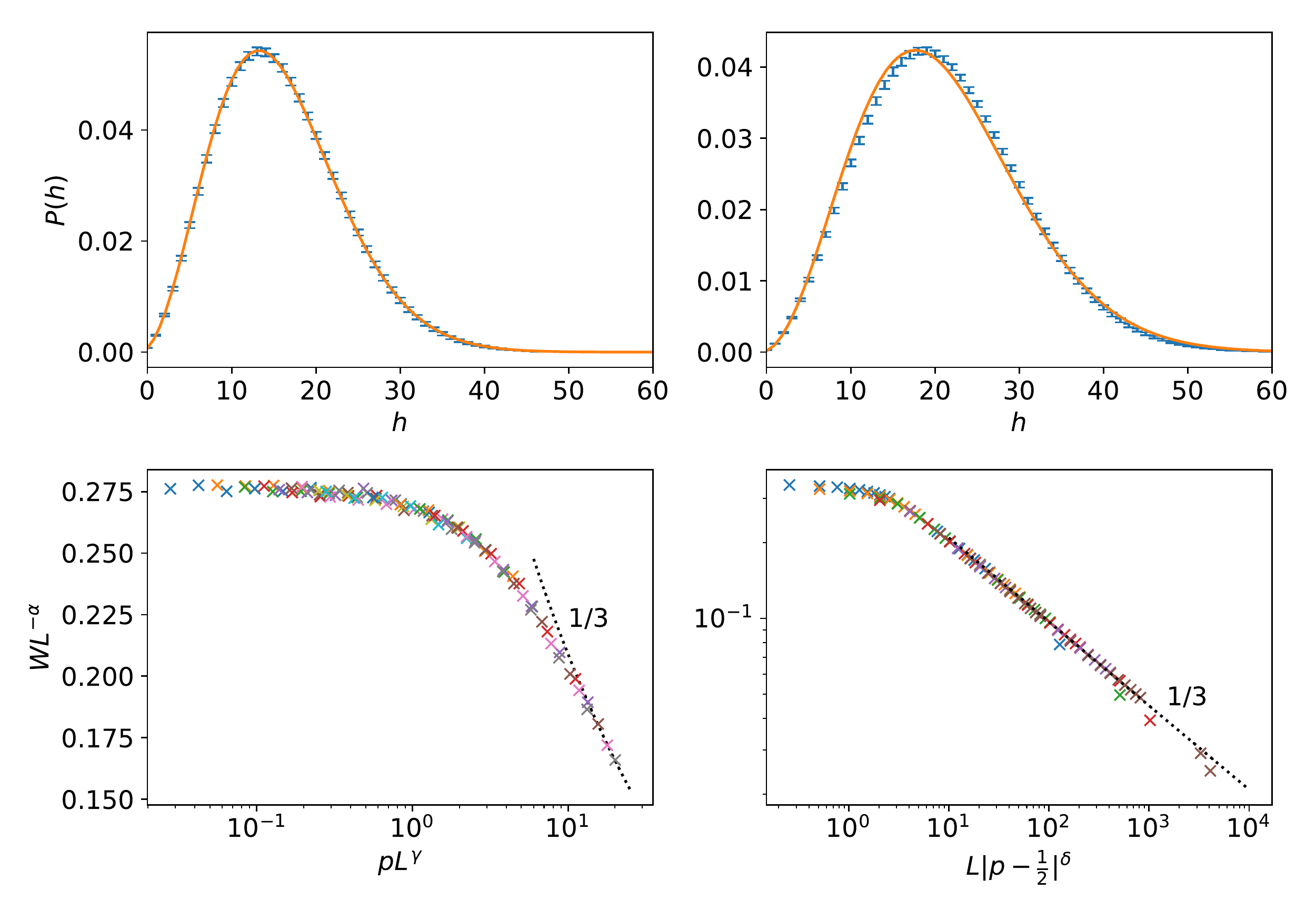}
 \caption{Top: Fit of $P(y)$ to Airy function solution (\ref{P(y)}) within RB phase. Left: $L=1024$, $p=0.5$, $u=0.4$ (near the exactly-solvable line). Right: $L=2048$, $p=0.25$, $u=0.25$ (far from the exactly-solvable line). Bottom: Crossover scaling functions. Left: Scaling function (\ref{co}) for width near $p=0$ at $u=0$ with $\alpha=\frac{1}{2}$ and crossover exponent $\gamma = 0.6$. Right: Scaling function (\ref{co2}) for width  for  $p\geq \frac{1}{2}$ at $u=0$ with $\alpha=\frac{1}{3}$ and crossover exponent $\delta = 1$.  In both lower panels, the dashed line corresponds to $x^{-1/3}$.}
\label{beast}
\end{figure}

\noindent{\em Roughness exponent in the RB phase.}---We now present our main numerical evidence that the roughness exponent $\alpha=\frac{1}{3}$ throughout the  RB phase. First, we find that the Airy function height distribution (\ref{P(y)}), with $\epsilon$ and $\beta$  free parameters, fits the numerically-determined distribution both near the exactly-solvable line and deep in the RB phase (see upper panels of Fig.~\ref{beast}).

Second, crossovers to known behavior are also consistent with $\alpha=\frac{1}{3}$. At $p=0$, the interface only grows downwards and cannot be affected by the membrane above it. Thus at $p=0$ the interface will exhibit the usual KPZ scaling with a roughness exponent $\alpha = \frac{1}{2}$. However, for $p>0$ in the RB phase, we expect $\alpha=\frac{1}{3}$ and  at finite $L$ there must be a crossover 
\begin{equation}
W \sim L^{\frac{1}{2}} g(pL^\gamma) 
\label{co}
\end{equation}
where $\gamma$ is a crossover exponent and $g(x)$ is a scaling function which tends to a constant for small $x$ and  has asymptotic behaviour   $g(x) \sim x^\phi$ for $x \gg 1$.  For  $\alpha$ to go from $\frac{1}{2}$ along $p=0$ to $\frac{1}{3}$ within the RB phase we require $\phi \gamma = -\frac{1}{6}$. The data collapse in Figure~\ref{beast} supports the scaling (\ref{co}) and  is consistent with $\phi = -\frac{1}{3}$ and $\gamma = \frac{1}{2}$.

The system should also cross over to  smooth phase  behaviour for $p>\frac{1}{2}$ at $u=0$. To study this we introduce a second crossover scaling function
\begin{equation}
W \sim L^{\frac{1}{3}} k\big(|p-{\textstyle\frac{1}{2}}|^\delta L\big) 
\label{co2}
\end{equation}
and find that for $p>\frac{1}{2}$ there is a scaling collapse with $\delta=1$. If the RB roughness exponent is $1/3$, as claimed, we expect $k(x) \sim x^{-\frac{1}{3}}$ in this regime, which is confirmed by Fig.~\ref{beast}. The scaling form (\ref{co2}) further implies a divergence of the width $W \sim  |p-\frac{1}{2}|^{-\frac{1}{3}}$ in the smooth phase as $p \searrow \frac{1}{2}$, similar to a nonequilibrium wetting transition \cite{Hinrichsen1997,Hinrichsen2003,Barato2010}.

\noindent{\em Dynamical arrest.}--- Finally, we return to the time dependence of the interfacial width, shown in Fig.~\ref{fig:dynscal}, which provides a physical picture of the origin of the roughness exponent $\alpha=\frac{1}{3}$. Taking the scaling form (\ref{width}), we can attempt to collapse $W(t)$ for different $L$ by choosing different combinations of $\alpha$ and $z$. The best collapse is obtained when we assume that the early time growth, $W\sim t^\beta$ is \emph{unaffected} by the membrane, and that the width saturates as $L^{\alpha}$ with $\alpha=\frac{1}{3}$. In the RB phase, the initial growth should be in the KPZ class $\beta=\frac{1}{3}$, which implies $z=\frac{\alpha}{\beta}=1$. On the transition line $p=\frac{1}{2}$, the nonlinearity in (\ref{KPZ}) is absent, and an Edwards-Wilkinson (EW) early-time growth $\beta=\frac{1}{4}$ is expected, implying $z=\frac{4}{3}$. The scaling collapse (Fig.~\ref{fig:dynscal}) is consistent with both expectations and suggests that the effect of the membrane is to prematurely \emph{arrest} the growth of the interface. The fact that different exponents are obtained for $p=\frac{1}{2}$ lends further credence to this being the true transition line.

\noindent{\em Conclusion.}--- In this paper we have introduced a simple model of an interface interacting with a membrane: the presence of the membrane obstructs the growth of the interface when they are in contact.
We have a determined a phase diagram that contains an unexpected phase RB in which the membrane is weakly bound to the interface but the roughness exponent of the interface is $\frac{1}{3}$ rather than the usual (for both KPZ and EW interfaces) value of $\frac{1}{2}$. Evidence that this is the true asymptotic  is provided by an exact solution along a special line (\ref{pwa}), where detailed balance holds, and various scaling analyses.

It would be of great interest to observe  the dynamical arrest  experimentally, for example, in the context of growing
bacterial colonies that motivated arrangement B. Such 
experimentally-relevant interfaces would usually be two-dimensional (2D), 
however we
expect similar phenomena to occur in 2D as for the one-dimensional case:
preliminary simulations \cite{Francesco} in 2D for $p=1$ indicate a
transition from a smooth to a rough phase  RA
with the same scaling (\ref{width}) for the width but with exponent
values corresponding to 2D KPZ $\alpha \simeq 0.4$, $\beta\simeq
0.2$. Of further interest would be to investigate the 2D analogue of
phase RB and associated exponents,  however simulation times for the 2D single-step model with $p<1$ are known to become prohibitive \cite{KK89}.

Many other open questions remain.  A more complete theory that
explains why the dynamical arrest occurs only in arrangement B, where
the membrane follows the interface from behind, is desirable. One
possibility is that the asymmetric form of the universal height
distribution of the growing KPZ interface \cite{Johansson,PS,SS}
implies that the interaction between interfacial peaks and troughs is
fundamentally different.  Finally, it should be noted that
experimental interfaces such as those we invoke will naturally exhibit
hydrodynamic couplings mediated by any fluid medium.  Such couplings
could generate effective long range interactions on the interface
which, it is known, can change scaling properties \cite{BS}.

\begin{acknowledgments}
The authors thank Bartek Waclaw for interesting discussions on growing bacterial colonies and Davide Marenduzzo for interesting discussions on Brownian ratchets. J.W.\ acknowledges support from EPSRC under a studentship
Grant No. EP/G03673X/1. J.W.\ and M.R.E.\ acknowledge the hospitality of the Weizmann Institute and M.R.E.\ acknowledges a Weston Visiting Professorship.
\end{acknowledgments}

\clearpage

\onecolumngrid
\appendix

\begin{center}
\large\bf Supplementary Figures
\end{center}

\renewcommand{\thefigure}{S\arabic{figure}}
\setcounter{figure}{0}

\begin{figure}[h!]
\includegraphics[width=0.6\linewidth]{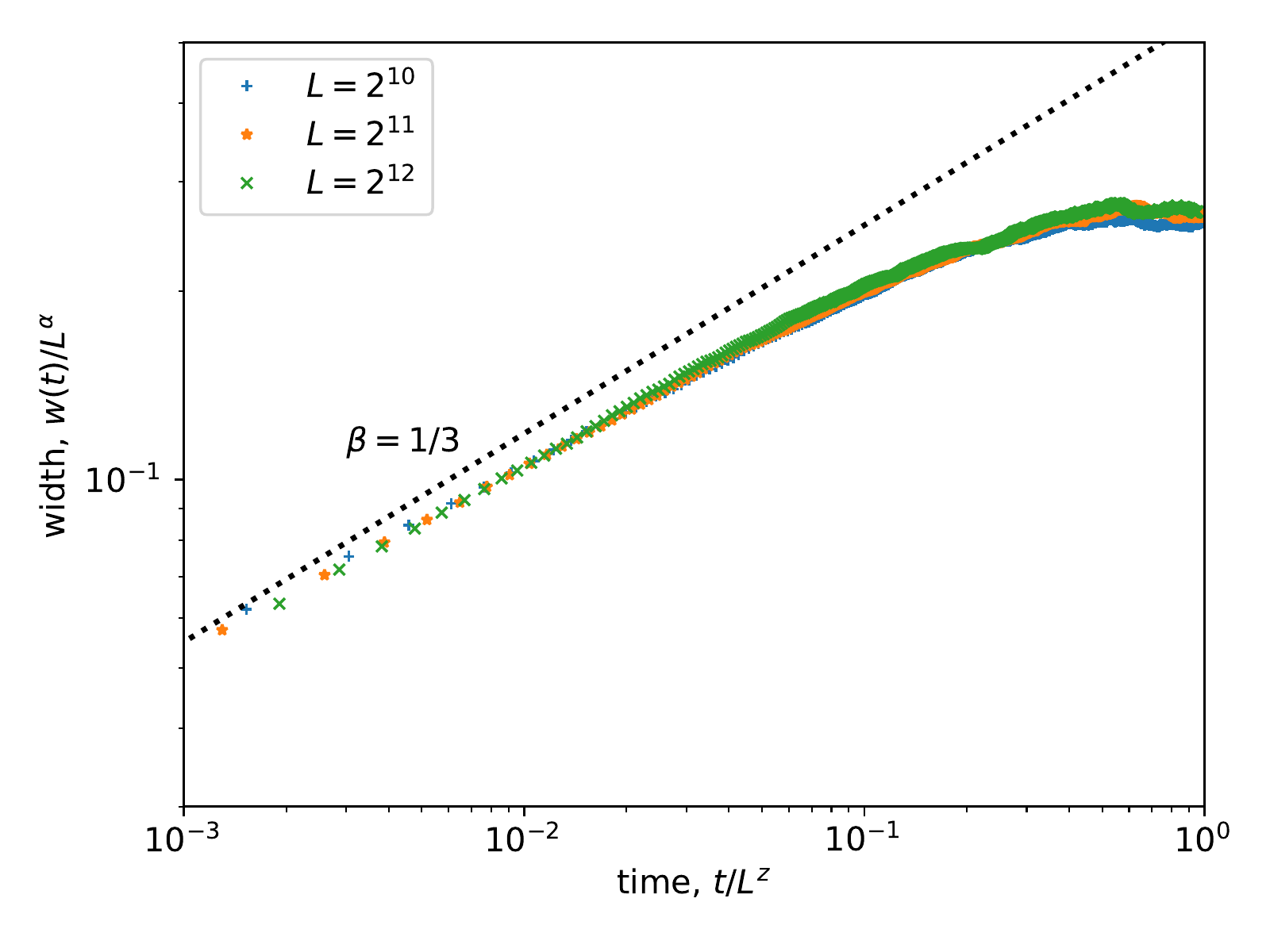}
\caption{Scaling of the interfacial width in the RA phase ($u=0.5, p=0.75$) with a KPZ roughness exponent $\alpha=\frac{1}{2}$ at different system sizes $L$. The early-time growth is consistent with the KPZ exponent of $\beta=\frac{1}{3}$.}
\end{figure}

\begin{figure}[h!]
\includegraphics[width=0.6\linewidth]{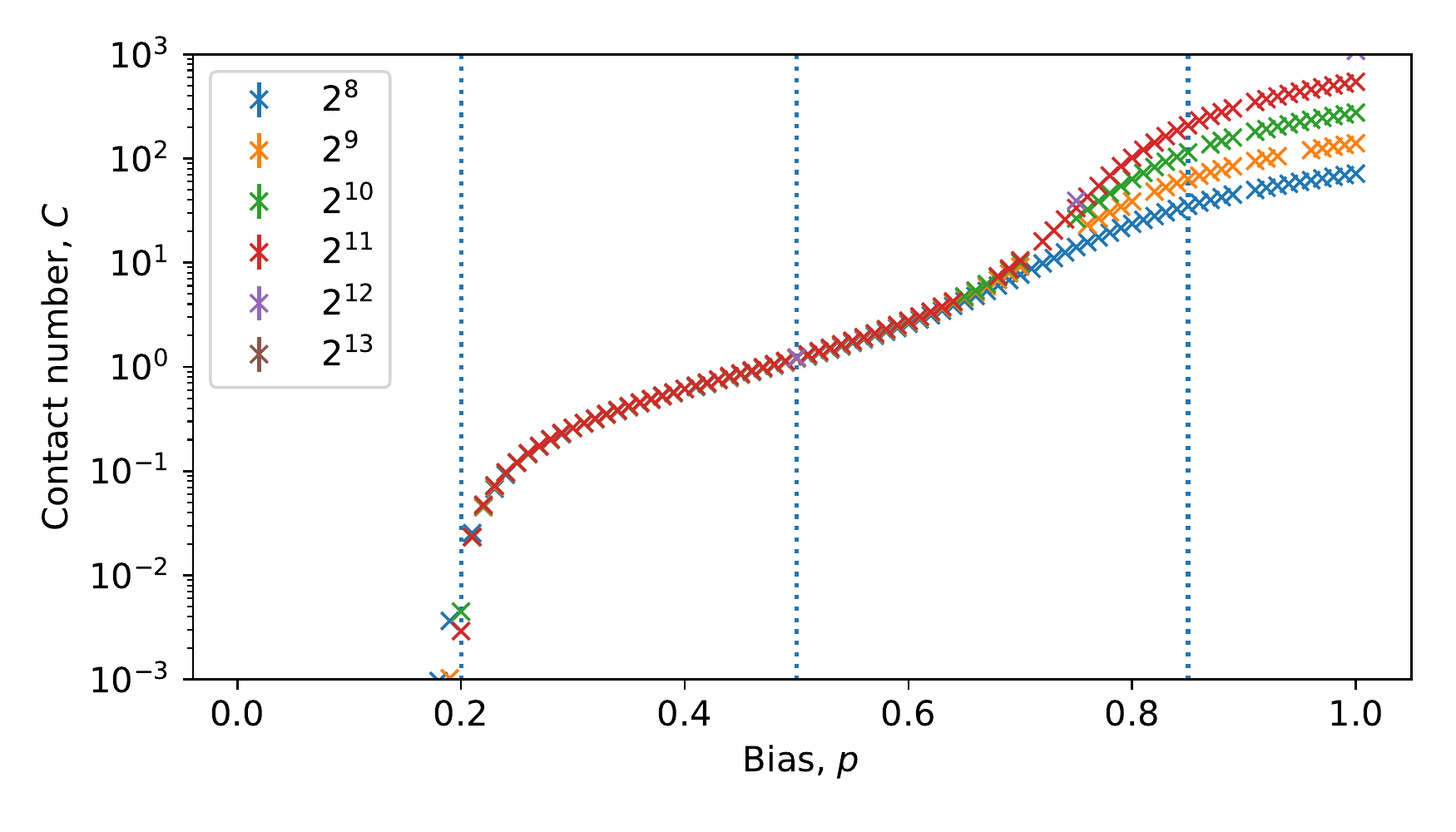}
\caption{Contact number $C$ against $p$ at $u=0.35$. For $p<p_1=2u-\frac{1}{2}=0.2$, the contact number falls to zero (the vertical axis is logarithmic), consistent with the unbound phase.  For $p\gtrsim p_2=u+\frac{1}{2}=0.85$ (the transition point may lie below $p_2$), the contact number increases extensively with the system size, consistent with a smooth interface that touches the membrane many times. In the intermediate range of $p$, $p_1 < p \lesssim p_2$, the contact count is independent of system size, consistent with a rough interface that contacts the membrane at only a few points.}
\end{figure}


\begin{thebibliography}{99}

\bibitem{HHZ}
Halpin-Healy T and Zhang  Y-C 
{\it Physics Reports} {\bf 254}, 215 (1995)

\bibitem{Krug97}
Krug J
{\it Advances in Physics} {\bf 46}, 139 (1997)

\bibitem{KK10}
Kriecherbauer T, Krug J 
{\it J. Phys. A: Math. Theor.} {\bf 43} 403001 (2010)

\bibitem{Takeuchi17}
Takeuchi KA
{\it Physica A} {\bf 504} 77 (2018)

\bibitem{KPZ} 
Kardar, M, Parisi, G, Zhang, Y-C
{\it Phys. Rev. Lett.}  {\bf 56} 889 (1986)

\bibitem{FV}
Family F, Vicsek T
{\it J. Phys. A} {\bf 18} L75 (1985)

\bibitem{KMHH}
Krug J, Meakin P, and Halpin-Healy T
{\em Phys. Rev. A} {\bf 45}  638 (1992)

\bibitem{Johansson}
Johansson K
{\it Commun. Math. Phys.} {\bf 209} 437 (2000)

\bibitem{PS}
Pr\"ahofer M, Spohn H
{\it Phys. Rev. Lett.} {\bf 84} 4882 (2000)

\bibitem{SS}
Sasamoto T,  Spohn H
{\it Phys. Rev. Lett.} {\bf 104} 230602 (2010)

\bibitem{TS}
Takeuchi KA, Sano M
{\it Phys. Rev. Lett.} {\bf 104} 230601 (2010)

\bibitem{TSSS}
Takeuchi KA, Sano M, Sasamoto T,  Spohn H
{\it Sci. Reports} {\bf 1} 34 (2011)

\bibitem{CFS17}
Chhita S,  Ferrari P L,  Spohn H 
{\em Ann. Appl. Prob} {\bf 28} 1573 (2018)

\bibitem{MS17}
Meerson B,  Schmidt J
{\it J. Stat. Mech.}  103207 (2017)

\bibitem{MKV16}
Meerson B, Katzav E, Vilenkin A
{\it Phys. Rev. Lett.}
{\bf 116} 070601 (2016)

\bibitem{LDMRS16}
 Le Doussal  P,  Majumdar S N,  Rosso A, Schehr G
{\it Phys. Rev. Lett.}  {\bf 117} 070403  (2016)

\bibitem{BS} 
Barabasi A-L, Stanley H E
{\it Fractal Concepts in Surface Growth} Cambridge University Press (1995)


\bibitem{LP}
Livi R and Politi P
{\it Nonequilibrium Statistical Physics} Cambridge University Press (2017)

\bibitem{KS88}
Krug J and Spohn H
{\it Phys. Rev. A}  {\bf 38}  4271 (1988)

\bibitem{MKHZ89} 
Medina E, Kardar M, Hwa T, Zhang, Y-C
{\it Phys. Rev. A}  {\bf 39}  3053 (1989)

\bibitem{DLSSa}
Derrida B, Lebowitz J, Speer ER, Spohn H
{\it Phys. Rev. Lett.} {\bf 67} 165 (1991)

\bibitem{DLSSb}
Derrida B, Lebowitz J, Speer ER, Spohn H
{\it J. Phys. A: Math. Gen} {\bf 24} 4805  (1991)

\bibitem{WT90}
Wolf DE, Tang L-H 
{\it Phys.  Rev.  Lett.} {\bf 65} 1591 (1990)

\bibitem{KM92}
Kandel D, Mukamel D
{\it EPL} {\bf 20} 325 (1992)


\bibitem{MRSB86}
Meakin P,   Ramanlal P,  Sander LM, Ball RC {\it Phys.
Rev.  A} {\bf 34} 5091 (1986)

\bibitem{KK89}
Kim JM  and  Kosterlitz JM 
{\it Phys. Rev. Lett.} {\bf 62} 2289 (1989)

\bibitem{Bray01}
Bray D 2001
{\em Cell Movements: From Molecules to Motility}
2nd edn (New York: Garland Publishing)

\bibitem{WBFEGM}
Wolff K,  Barrett-Freeman C,  Evans M R,  Goryachev A B and  Marenduzzo D
{\em Physical Biology}, {\bf 11}, 016005 (2014)

\bibitem{WSG17}
Wang X, Stone H. A., Golestanian R. 
{\em New Journal of Physics}, {\bf 19} 125007 (2017)

\bibitem{BWpc}
Waclaw B, private communication.

\bibitem{POO93} Peskin C S, Odell G M, Oster G F, {\it Biophysical Journal} {\bf 65}, 316 (1993)

\bibitem{BE2001}
Blythe RA, Evans MR
{\it Phys. Rev. E} {\bf 64} 051101 (2001)

\bibitem{Hinrichsen1997}
Hinrichsen H, Livi R, Mukamel D, Politi A
{\it Phys. Rev. Lett.} {\bf 79}, 2710 (1997)

\bibitem{Hinrichsen2003}
Hinrichsen H, Livi R, Mukamel D, Politi A
{\it Phys. Rev. E} {\bf 68}, 041606     (2003)

\bibitem{Barato2010} Barato A C {\it J. Stat. Phys}
{\bf 138}  728 (2010)

\bibitem{Francesco} Cagnetta F {\em private communication}


\end{thebibliography}
\end{document}